\documentclass[12pt]{article}
\newcommand{\ra}{\rightarrow}

\newcommand{\no}{\nonumber}
\newcommand{\be}{\begin{equation}}
\newcommand{\ee}{\end{equation}}
\newcommand{\ba}{\begin{eqnarray}}
\newcommand{\ea}{\end{eqnarray}}

\newcommand{\dta}{\mbox{$\delta$}}
\newcommand{\fee}{\mbox{$\varphi$}}
\newcommand{\al}{\mbox{$\alpha$}}

\newcommand{\eps}{\mbox{$\epsilon$}}

\newtheorem{theo}{Theorem}[section]

\newtheorem{prop}{Proposition}[section]

\begin{document}
\title{Statistical Analysis of a \\
Semilinear Hyperbolic System Advected by a \\
White in Time Random Velocity Field}
\author{Gregory Eyink \thanks{
Dept of Math, Univ of Arizona, Tucson, 
AZ 85721. Email: eyink@math.arizona.edu.} and 
Jack Xin\thanks{Dept of Math \& TICAM,
Univ of Texas at Austin, Austin, TX 78712. 
Email: jxin@math.utexas.edu. }
}
\date{ }
\baselineskip=18pt
\maketitle
\begin{abstract}
We study a system of semilinear hyperbolic equations passively 
advected by smooth white noise in time random velocity fields. 
Such a system arises in modeling non-premixed isothermal 
turbulent flames 
under single-step kinetics of fuel and oxidizer. 
We derive closed equations for one-point and multi-point 
probability distribution functions (PDFs) 
and closed form analytical formulas for the one point PDF function,  
as well as the two-point PDF function under homogeneity and isotropy.  
Exact solution formulas allows us to 
analyze the ensemble averaged fuel/oxidizer concentrations 
and the motion of their level curves. 
We recover the empirical  
formulas of combustion 
in the thin reaction zone limit and show that 
these approximate formulas can either underestimate or 
overestimate average concentrations when reaction zone is not tending to 
zero. 
We show that 
the averaged reaction 
rate slows down locally in space due to 
random advection induced diffusion; and that  
the level curves of ensemble averaged concentration undergo  
diffusion about mean locations. 

\end{abstract}
\vspace{.1 in}

\thispagestyle{empty}
\newpage
\setcounter{page}{1}
\section{Introduction}
We are interested in statistical properties of solutions of 
the following passively advective semilinear hyperbolic systems:
\be
u_{t} + v(t,x,\omega)\circ \nabla_{x} u = f(u), \label{sem1}
\ee
where $u \in R^m$, $m \geq 1$; $x \in R^N$, $N\geq 2$; $f: R^m \ra R^m$, 
a smooth nonlinear map; $v(t,x,\omega)= \bar{v}(x) + v'(x,t,\omega)$, 
where $\bar{v}$ is the mean vector field (e.g. a constant vector field), 
$v'$ the fluctuation random field,  stationary
Gaussian in space and white in time, with mean zero, and covariance: 
\be
<v'_{i}(\vec{x},t)v'_{j}(\vec{x}',t')> = D_{ij}
(\vec{x} -\vec{x}')\delta(t-t'),
\label{sem1a}
\ee
where $D=(D_{ij})(\cdot)$ is smooth. The $\circ$ denotes 
the Stratonovitch sense of stochastic product. The coupling 
of different components of $u$ is through the lower order term $f$. The 
specific forms of $f$ in applications 
often permit invariant regions hence uniform maximum norm bounds exist 
on the solutions for all times. 

An example of (\ref{sem1}) is the following system:
\ba
Y_{F,t} + v(t,x,\omega)\circ \nabla_{x}Y_F =\kappa \Delta Y_{F} - K Y_F Y_O, 
\no \\
Y_{O,t}+ v(t,x,\omega)\circ \nabla_{x}Y_O =\kappa \Delta Y_{O} - K Y_F Y_O,
\label{sem2}
\ea
where $\kappa \geq 0$ is diffusion constant, $K > 0$ reaction rate constant,
$Y_F$ ($Y_O$) is the concentration of fuel (oxidizer), $v$ a prescribed 
random turbulent velocity field. 
Such a system arises in isothermal 
non-premixed turbulent combustion under single step 
reaction: $Y + \nu O_x \rightarrow P$, where $P$ is product, $\nu$ 
stoichiometric constant, see \cite{Bor} (chapters 2 and 6), 
\cite{LW} (chapters 3 and 5), 
and 
\cite{Maj00}. We shall be concerned with the inviscid regime where $\kappa =0$. 
The common quantities of interest include the ensemble averaged 
concentrations and reaction rate $E[Y_F]$, $E[Y_O]$, $E[Y_FY_O]$ 
(see e.g. \cite{Bor}, section 6.7, \cite{LW} chapter 5). 

It is known that these moments of solutions as well as the underlying  
probability distribution functions (PDF's) do not satisfy any 
closed equations in general. To proceed with an exact analytical treatment, 
we make the white noise in time assumption \'a la 
Kraichnan on the velocity field \cite{Kr68}, \cite{Kr74}. 
Additionally, the sample paths 
of velocity field are no less smooth than Lipschitz continuity, see  
the recent work \cite{Son99}, \cite{Bal99} in this so called Batchelor limit. 
This permits us to 
get closed equations for PDFs for nonlinear systems of equations such 
as (\ref{sem1}). 
To understand the analytical structures 
of the class of stochastic PDE problems like (\ref{sem1}) 
and make progress without 
resorting to closure approximations, the white in time assumption offers 
a unique starting point, and helps one to gain insight. 
This line of inquiry 
has been pursued in many recent works
\cite{Maj93}, \cite{Cons98}, \cite{BGK}, among others. 
For successes of the white noise model in the context of 
the linear passive scalar model, see the in-depth 
survey by Majda and Kramer \cite{Maj99}. For an extension to 
finite time correlation from the white noise in time 
linear passive scalar model, see 
the recent work \cite{Bal99}. 

We mention in passing that 
 similar semilinear systems also arise in bioremediation and transport 
in porous media \cite{Gelhar}, \cite{OVa}. In \cite{XH}, system (\ref{sem2}) 
was derived in a fast reaction limit of microbial reaction when the 
substrate retardation factor is equal to one. However, 
the velocity $v$ in that particular application is 
spatially random and independent of time \cite{Gelhar}. 
The random dependence of velocity field in time is more
relevant in turbulent combustion.  
A model similar to (\ref{sem2}), 
was recently considered by Majda and Souganidis \cite{Maj00} for 
the case of two-scale periodic velocity field (oscillatory in 
both space and time) to understand the large 
scale effect of turbulent mixing in non-premixed combustion and 
condensation-evaporation in cloud physics. Mean field 
equations were obtained in the limit of large scale separation 
(homogenization limit). Here we obtain the exact equations without 
scale separation in space but with rapidly decorrelated velocity 
in time. 

Within the framework of the white noise model, we are able to 
analyze several important quantities (
ensemble average of solutions and reaction rate) often encountered in practice. 
We also use our exact results to understand approximate formulas 
in combustion literature on calculations of these averaged solutions. 
Moreover, we have made rigorous previous formal derivations of PDF equations 
by Kraichnan and coworkers \cite{Kr74}, \cite{CCK89}, 
\cite{Kr93}. 

In section 2, we derive the one point PDF 
($P=P(u,x,t)$) equation and exact solution formulas  
on averages of solutions under the initial condition 
$P(u,x,0^+) =\delta_{u=u_0(x)}$, where $u_0$ is 
a bounded nonnegative determinisitic (vector) function such that 
the reaction $f(u)$ is integrable. 
In the thin reaction limit of (\ref{sem2}), 
or $f(u)=0$ except on a low dimensional 
manifold, we recover the empirical formulas in combustion (\cite{Bor} 
section 6.7), and show how (under)overestimation can occur with 
these formulas when 
reaction zone is not thin. 
Moreover, the average reaction rate $E[f(u)]$ 
decreases pointwise in time to zero with a much faster rate than the 
advection free case due to  
enhanced diffusion caused by random advection. With both a scalar model 
of Fisher type ($f(u)=u(1-u)$) 
and the model system (\ref{sem2}), we show that the 
level curves of averaged solutions undergo diffusion about the mean location.

In section 3, we derive the 
multipoint PDF equation and analyze its solutions. By a comparison 
argument, we show 
the decay of the correlation functions for
large times using the correlation functions of Kraichnan's passive 
scalar model. We also find the two-point PDF solution formula in closed 
form under isotropy and homogeneity conditions. 

Section 4 are concluding remarks.  
Many open issues remain for future work. 
One of them is to extend results here to the regime where  
velocity 
time correlation is a finite value away from zero. The other 
is to study more properties of level curves of concentration 
variables using PDF information. For results along this line using 
a different approach, see Constantin and Procaccia \cite{Cons94}.

\section{One Point PDF Equation and its Solutions}
\setcounter{equation}{0}
\subsection{Derivation of One Point PDF Equation}

Let us first recall a representation of Gaussian random velocity $v$, 
so called proper orthogonal decomposition, 
see \cite{JanRai} among others. The 
velocity field $v(t,x,\omega)$ is formally the time derivative of 
a cylindrical Brownian motion $W_t$ on a separable Hilbert space $H$, 
or a Gaussian process indexed by $H\times R^{+}$ with covariance matrix 
$cov(W_t(h),W_s(h'))=\min (t,s) <h,h'>_{C}$, $h$, $h'\in H$. The underlying 
space $H$ carries information of the original $x$ variable. 
The covariance of velocity 
$v$ can be regarded as a positive definite bilinear operator acting on $H$.
If $D=D(x,y)$ is in $L^2_{loc}(dx \times dy)$, then
there exists an orthogonal basis $e^{(n)}(x) \in R^N$ of space $H$ 
such that: 
\be
D_{i,j}(x,y) =\sum_{n}\, e_{i}^{(n)}(x) e_{j}^{(n)}(y), \;\; 1\leq\,  
i,\, j, \, \leq N. \label{D1}
\ee
Now let $W^{(n)}(t,\omega)= W_{t}(e^{(n)})$, 
which is a sequence of 
independent Wiener processes (with $\omega$ omitted below), 
the representation reads:
\be
v(x,t,\omega) = \sum_{n} e^{(n)}(x) \, dW^{(n)}(t)/dt. \label{D2}
\ee

With (\ref{D2}), the original system (\ref{sem1}) can be put into:
\be
d_{t} u = -\sum_{n}\, (e^{(n)}(x)\cdot \nabla_{x} u )\circ dW^{(n)}(t) 
+ f(u) dt \equiv -\sum_{n}\, g^{(n)}(x,u) \circ dW^{(n)}(t) 
+f(u) dt, \label{D3}
\ee
where $\circ$ refers to the Stratonovich sense of differential product. 
The equation (\ref{D3}) with Lipschitz velocity $u$ falls into the 
category of stochastic flows studied in Kunita \cite{Kunita}, solutions 
exist and are unique, and can be approximated by finite dimensional 
stochastic ODE's. Recall a well-known conversion from Stratonovich 
to Ito (nonanticipating) stochastic ODE's ($X=X(t,\omega) \in R^M$, 
$g^{(m)} \in R^M$):
\be
d X = f(X) dt + \sum _{m=1}^{M}\, g^{(m)}(X)\circ dW^{(m)}(t), \label{D4}
\ee
where $W^{(m)}(t)$'s are independent Wiener processes, $f$ and $g$ 
are Lipschitz functions of $X$. Then the equivalent Ito equation is:
\be
d X = [ f(X) + {1\over 2}\sum_{m=1}^{M}\, 
g^{(m)}(X)\cdot \nabla_{X}\, g^{(m)}(X)] dt
+ \sum_{m=1}^M g^{(m)}(X) dW^{(m)}(t), \label{D5}
\ee
the new drift term is the so called noise induced drift. A similar 
conversion for (\ref{D3}) generates the drift term:
\ba
{1\over 2}\sum_{n}\, \int \, dy\, g^{(n)}(y,u) \nabla_{u(y)} 
g^{(n)}(x,u)  
& =  & {1\over 2}\sum_{n}\, \int \, dy\, e^{(n)}(y)\cdot\nabla_y 
u(y)\,\,\, e^{(n)}(x) \cdot \nabla_{x} \delta (x-y) \no \\
& = & {1\over 2}\nabla_x\cdot \int \, dy\, [\sum_{n}\,e^{(n)}(x) 
\otimes e^{(n)}(y)] \cdot \nabla_{y}u(y) \delta (x-y) \no \\
& = & {1\over 2} \, \nabla_{x}\cdot \int \, dy D(x,y)\nabla_{y} u(y)
\delta (x-y) \,\,\,\,{\rm using}\,\,\,\,{\rm (2.1)}\no \\
& = & {1\over 2}\nabla_{x}\cdot ( D(x,x)\nabla_{x} u(x))
\no \\
& = 
& {1\over 2}\nabla_{x}\cdot (D(0)\nabla_{x} u(x)). \label{D6}
\ea
It follows that the Ito form of the equation is:
\be
d_{t} u = -\sum_{n}\, (e^{(n)}(x)\cdot \nabla_{x} u ) dW^{(n)}(t) 
+ [f(u) +  {1\over 2}\nabla_{x}\cdot (D(0)\nabla_{x} u)] dt, \label{D7}
\ee
or in PDE form:
\be
u_{t} + v\cdot \nabla_x u = f(u) +  
{1\over 2}\nabla_{x}\cdot (D(0)\nabla_{x} u), \label{D8}
\ee
The advantage of the Ito form (\ref{D7}) or (\ref{D8}) is that 
the stochastic advection term has mean zero.

Let $\fee=\fee(u): R^N \ra R^N$ be a smooth nonlinear map, and 
$J_{\fee}=J_{\fee}(u)$ be its Jacobian. Multiplying system (\ref{sem1}) on the 
left by $J_{\fee}$, we find:
\be
(\fee(u))_{t} + v\circ \nabla_{x} \fee(u) = J_{\fee}(u)f(u), \label{D9}
\ee
whose equivalent Ito system of equations are with the same derivation as 
above:
\be
\fee(u)_{t} + v \cdot  \nabla_{x} \fee(u) = 
{1\over 2}\nabla_x \cdot (D(0) \nabla_x \fee(u)) + J_{\fee}(u) f(u). \label{D10}
\ee
Taking ensemble mean of (\ref{D10}), we get:
\be
E[\fee(u)]_{t} =  {1\over 2}\nabla_x \cdot (D(0) \nabla_x E[\fee(u)]) 
+ E[J_{\fee}(u) f(u)], \label{D11}
\ee
and the last term is equal to 
$- \int \fee(u) \nabla_{u}\cdot (f(u)P(u,x,t))\, du$. Hence it follows 
from (\ref{D11}) that the $P(u,x,t)$ satisfies the equation:
\be
P_{t} + \nabla_{u}\cdot (f(u)P) = {1\over 2}\nabla_x \cdot (D(0) \nabla_x P).
\label{D12}
\ee

We may summarize the results of this section in the following:
\begin{theo}
Let $P(u,x,t)$ be the one-point probability density function of the 
stochastic solution $u(x,t)$ of the semilinear hyperbolic system
(\ref{sem1}) with white noise in time, spatially Lipschitz, incompressible,
Gaussian random velocity field $v(x,t)$. Then $P$ solves the closed 
equation (\ref{D12}).
\end{theo}

\subsection{Applications}
\subsubsection{A Model Equation with Fisher Nonlinearity}
Let us first consider a scalar model equation of the form (\ref{sem1}) with 
$f(u)=u(1-u)$, the so called Fisher (or KPP, Kolmogorov-Petrovsky-Piskunov) 
nonlinearity. We shall assume that the advection veocity has mean equal to 
zero. If the advection is identically zero, then the solution is:
\be
u(x,t)={u_{0}(x) e^{t}\over (1-u_{0}(x)) +u_{0}(x) e^{t}}, \label{F1}
\ee
which is between zero and one if the initial data $u_{0}(x)$ is so. 
As $t \ra +\infty$, $u \ra 1$ if $x \in \{x: u_{0}(x) > 0\}$, 
and $u \ra 0$ if $x \in \{x: u_{0}(x) =0\}$. In particular, any 
characteristic function is invariant in time. For the random advection, 
the PDF equation (\ref{D12}) reads:
\be
P_{t} +(u(1-u)P)_{u}= {1\over 2}D_0 \Delta_{x} P, \label{F2}
\ee
with initial data $P_{0}=P_{0}(u,x)=(4\pi \sigma_u^{2})^{-N/2}
\exp\{-(u -u_0(x))^2/(4\sigma_u^{2})\}$, 
where $u_0(x)$ is a bounded measurable 
function between zero and one, and $\sigma_{u}$ is a small parameter.
Such initial data is convenient for examining deterministic 
initial data by taking $\sigma_{u} \ra 0$ limit after we find the 
general formula of the PDF.

Let us solve (\ref{F2}) by taking Fourier transform in $x$ (
justified by a suitable truncation, see next subsection for details) to get:
\be
\hat{P}_{t} + u(1-u)\hat{P}_{u} =(2u -1 -D_0|\xi|^{2}/2)\hat{P}, 
\label{F3}
\ee
which is then integrated by the method of characteristics. The characteristic 
curve is:
\be
u ={u_{0}e^{t}\over (1-u_0)+u_0 e^{t}}, \label{F4}
\ee
whose inverse is:
\be
u_0= {u \over u +(1-u)e^{t}}. \label{F5}
\ee
So:
\ba
\hat{P} & = & \hat{P}_0(u_0,\xi)\exp\{2 \int_{0}^{t} {u_0 e^{t'}\over 
(1-u_0)+u_{0}e^{t'}}\, dt' - t -{D_{0}\over 2}|\xi|^{2} t \} \no \\
& = & \hat{P}_0(u_0,\xi) (1 -u_0 +u_0 e^{t})^{2}
\exp\{ -t -{D_{0}\over 2}|\xi|^{2} t \}  \no \\
& = & \hat{P}_{0}\left ({u \over u +(1-u)e^{t}}, \xi \right )
{e^{t -D_{0}|\xi|^{2}t/2}\over (u +(1-u)e^{t})^{2}}. \label{F6}
\ea
It follows by taking inverse Fourier transform in $\xi$ that:
\be
P(u,x,t)={e^t (2\pi D_0 t)^{-N/2} \over (u+(1-u)e^{t})^{2}}\int_{R^N}\, dy \, 
e^{-|x-y|^{2}/(2D_0 t)} (4\pi\sigma_u^{2})^{-N/2} 
e^{-\left ( {u\over u+(1-u)e^{t}} -u_{0}(y)\right )^{2}/ (4\sigma_u^{2})}. 
\label{F7}
\ee

Now we calculate all moments of solution ($f=f(u)=u^p$, $p \geq 1$):
\ba
& & E[f(u)](x,t)  =  \int\, du\, f(u) P(u,x,t) \no \\
& = &  
\int\, dy\, K(t,x,y) \int\, du \, {e^{t}f(u)\over (u+(1-u)e^{t})^{2}}
\,  (4\pi\sigma_u^{2})^{N/2} 
e^{-\left ( {u\over u+(1-u)e^{t}} -u_{0}(y)\right )^{2}
/ (4\sigma_u^{2})}, \label{F8}
\ea
where $K$ is the heat kernel with diffusion constant $D_0/2$. The inner 
integral in (\ref{F8}) is equal to (via the change of variable 
$v= u/(u +(1-u)e^{t})$):
\be
\int\, dv\, (4\pi\sigma_u^{2})^{-N/2} e^{-(v -u_{0}(y))^{2}/(4\sigma_u^{2})}
f(ve^{t}/(1-v +ve^{t})). \label{F9}
\ee
Taking the limit $\sigma_{u} \ra 0$, we find that (\ref{F9}) 
converges to $f(u_0(y) e^t /(1 -u_0(y) +u_0(y) e^t))$, and:
\be
E[f(u)] \ra \int_{R^N}\, dy\, K(t,x,y) f\left ({u_{0}(y) e^{t} \over 
1- u_0(y) + u_{0}(y) e^{t}}\right ). \label{F10}
\ee

In particular, if the initial data is a front, namely, 
$u_0(y)= \chi(\{x: x_{1} \in [0,\infty) \}=\chi(R^N_{+})$, 
(\ref{F10}) reduces to:
\be
\int_{R^N_{+}}\, dy\, K(t,x,y)
 =  
(2\pi D_0 t)^{-1/2} \int_{0}^{\infty}\, dy_1\, e^{-(x_1 -y_1)^{2}/(2D_0 t)} 
=  \pi^{-1/2}\int_{{x_{1}\over \sqrt{2D_0 t}}}^{\infty}\, e^{-z^2}\, dz. 
\label{F11}
\ee
To probe the front motion, we look at the level set 
$X(t)$ such that $E[f(u)](t,X(t)) = c \in (0,1)$. For large $t$, 
(\ref{F11}) shows that:
\be
X(t) = (X_1(t), X_2, \cdots, X_N) \sim (z_0\sqrt{2D_0 t},X_2,\cdots,X_N), 
\label{F12}
\ee
where $z_0$ is the unique number so that 
$\int_{z_0}^{\infty} e^{-z^2} dz = \pi^{1/2} c$. The number 
$z_0 > 0 (<0)$ if $c > 1/2 (< 1/2)$, $z_0=0$ if $c=1/2$. This implies that 
the random front in the average sense undergoes normal diffusion about 
its mean, in this case $x_1=0$. 
Of course, the random level set $\{x: u(t,x)=c \in (0,1)\}$ 
is more complicated and analysis of its almost sure behavior 
requires more information (multipoint statistics).  

\subsubsection{The Model Combustion System}

Let us consider the one-point statistics of 
the two by two combustion system (\ref{sem2}) with $\kappa=0$, $K=1$,  
by studying the initial value problem of (\ref{D12}) with 
$f(u) =u_1 u_{2}(-1,-1)$, and initial data:
\[ P(u,x,0)={1\over (4\pi \sigma_u^{2})^{N/2}}
\exp\{-{1\over 4\sigma_u^{2}} (( u_1 - u_1^0(x))^2
+(u_2 - u_2^0(x))^2)\}, \]
the smoothed delta function located at 
$(u_1^0,u_2^0)(x)$. We shall find a solution formula then take the limit 
$\sigma_u \ra 0$. To use Fourier transform in $x$, we need to truncate 
$P(u,x,0)$ at large $x$, this can be done by multiplying to 
$P(u,x,0)$ a smooth function $\psi_R(x)$ compactly supported 
in the ball $B_R \in R^N$. The truncated data is denoted by 
$P_R(u,x,0)$. Fourier transforming (\ref{D12}) in $x$ gives:
\be
\hat{P}_{t} -(u_1 u_2 \hat{P})_{u_1} - (u_1 u_2 \hat{P})_{u_2} = 
-D_0|\xi|^2\hat{P}/2, \label{D13}
\ee
with initial data $\hat{P}_{R}(u,x,0)$, and $D_0=D(0)$. 
Equation (\ref{D13}) is put to the 
form:
\be
\hat{P}_{t} -u_{1}u_{2}\hat{P}_{u_1} -u_1 u_2\hat{P}_{u_2} = 
(u_1 +u_2 -D_0|\xi|^2/2)\hat{P}, \label{D13a}
\ee
which we solve by the method of characteristics. The characteristic $\Gamma$ 
equations: $u_{i,t}=-u_1\, u_2$, $i=1,2$, give solutions:
\be
u_1 = {c_0 c_1 e^{c_0 t}\over 1 + c_1 e^{c_0 t}},\;\; 
u_2 = - {c_0 \over 1+ c_1 e^{c_0 t}}, \label{D14}
\ee
and useful relations:
\be
{u_1 \over u_2} = - c_{1}e^{c_0 t}, \; c_0 = u_1 - u_2, \; 
c_1 = - {u_1\over u_2}e^{(u_2 - u_1)t}. \label{D15}
\ee
Along $\Gamma$, $\hat{P}$ obeys in view of (\ref{D13a}) and (\ref{D14}):
\[ \hat{P}_{t}= (u_1 + u_2 -D_0 |\xi|^2/2)\hat{P} = 
\left ({2c_0 c_1 e^{c_0 t}\over 1 + c_1 e^{c_0 t}} - c_0 -D_0|\xi|^2/2 
\right )\hat{P}, \]
and so:
\be
\hat{P}(\xi,t) = \hat{P}_R\left (u_1={c_0 c_1\over 1+c_1},
u_2={-c_0\over 1+c_1},\xi\right ){ (1+c_1 e^{c_0 t})^{2} \over 
(1+c_1)^2}e^{-(c_0 +D_0|\xi|^2 )t}, \label{D16}
\ee
Using (\ref{D15}) to write (\ref{D16}), we find:
\ba
\hat{P}(u,\xi,t) &=& \hat{P}_R\left ( { (-u_1 +u_2)\, u_1\, e^{(u_2 -u_1)t}
\over u_2 - u_1 e^{(u_2 -u_1)t}}, 
{(u_2 -u_1)u_2 \over u_2 - u_1 e^{(u_2 -u_1)t}}, \xi \right )\, \no \\
& & \cdot {(u_2 -u_1)^{2}\over (u_2 -u_1 e^{(u_2 -u_1) t})^2} 
e^{-(u_1 -u_2 +D_0|\xi|^2/2)t},
 \label{D17}
\ea
where:
\be
 \hat{P}_{R}(u,\xi) = \int \, dx\, \psi_{R}(x)\, 
{1\over (4\pi\sigma_{u}^{2})^{N/2}} 
\exp\{-(u_1 -u_{1}^{0}(x))^2/4\sigma_{u}^{2}
-(u_2 -u_{2}^{0}(x))^2/4\sigma_{u}^{2} - i\xi \cdot x \}.
\label{D18}
\ee 
Substituting (\ref{D18}) into (\ref{D17}), and taking inverse Fourier 
transform, we have:
\ba
P(u,x,t) & = & (2\pi)^{-N} \int_{R^N}\, dy\, \psi_{R}(y)  
{1\over (4\pi\sigma_u^{2})^{N/2}} \exp\{-\left ( 
{ (-u_1 +u_2)\, u_1\, e^{(u_2 -u_1)t}
\over u_2 - u_1 e^{(u_2 -u_1)t}} 
 -u_{1}^{0}(y)\right )^2/4\sigma_u^{2}  \no \\
& & -\left ({(u_2 -u_1)u_2 \over u_2 - 
u_1 e^{(u_2 -u_1)t}}-u_{2}^{0}(y)\right )^2/4\sigma_u^{2} 
\} \no \\
& & \cdot 
{(u_2 -u_1)^{2}\over (u_2 -u_1 e^{(u_2 -u_1) t})^2} 
e^{(u_2 -u_1)t} \int \, d\xi\, e^{-D_0 |\xi|^2 t/2 - i\xi\cdot y + i\xi\cdot x},
\no
\ea
where the inner integral gives the heat kernel: $(2\pi D_0 t)^{-N/2}\exp\{ 
-|x-y|^2/(2D_0 t)\}$. The exponential decay in $y$ allows us to 
remove the truncation $\psi_R$ by letting $R\ra +\infty$. It follows that:
\ba
P(u,x,t) & = & {(u_2 -u_1)^{2}\over (u_2 -u_1 e^{(u_2 -u_1) t})^2} 
e^{(u_2 -u_1)t} \int_{R^N}\, dy\, (2\pi D_0 t)^{-N/2}\exp\{ 
-|x-y|^2/(2D_0 t)\} \no \\
& & \cdot {1\over (4\pi\sigma_u^{2})^{N/2}} 
\exp\{-\left ( { (-u_1 +u_2)\, u_1\, e^{(u_2 -u_1)t}
\over u_2 - u_1 e^{(u_2 -u_1)t}}  -u_{1}^{0}(y)\right )^2/4\sigma_u^{2} \no \\
& & -\left ({(u_2 -u_1)u_2 \over u_2 - 
u_1 e^{(u_2 -u_1)t}}-u_{2}^{0}(y)\right )^2/4\sigma_u^{2} 
\}. \label{D19}
\ea
Note that the denominator $u_2 - u_1\, e^{(u_2 -u_1)t}$ being zero is 
not a singularity due to a similar term in the exponential of the 
Gaussian. 

The PDF formula (\ref{D19}) can be further simplified by taking 
$\sigma_u \ra 0$ for any finite time $t$. This is most conveniently done when 
we calculate the average quantities, $E[u_1\, u_2]$, $E[u_1]$, $E[u_2]$, 
 in the limit $\sigma_{u} \ra 0$.
\ba
& & E[u_1\, u_2]  = \int u_1\, u_2 P(u,x,t)\, du_1\, du_2 
\no \\
& \ra & \int \, dy\, (2\pi D_0 t)^{-N/2}\exp\{ 
-|x-y|^2/(2D_0 t)\}\, u_{1}^{0}(y)\, u_{2}^{0}(y)\, J(u_{1}^{0},u_{2}^{0},t),
\label{D19a} 
\ea
where $J=J(v_1,v_2,t)= \, det \,{\partial(u_1,u_2)\over \partial (v_1,v_2)}$, 
and the mapping $(v_1,v_2) \ra (u_1,u_2)$ at any time $t$ is the inverse of:
\be
v_{1}={ (-u_1 +u_2)\, u_1\, e^{(u_2 -u_1)t}
\over u_2 - u_1 e^{(u_2 -u_1)t}},\; 
v_{2}= {(u_2 -u_1)u_2 \over u_2 - 
u_1 e^{(u_2 -u_1)t}}. \label{D19b}
\ee
Notice that $(v_{1},v_{2})$ of (\ref{D19b}) satisfies the 
system: $v_{i,t}= v_{1}v_{2}$, $i=1,2$, with initial data 
$(v_1,v_2)(0)=(u_1,u_2)$. So the inverse map is the solution of 
the system: $u_{i,t}=-u_{1}u_{2}$ with initial data $(v_1,v_2)$, or:
\be
u_{1}={(v_1 -v_2)v_1 e^{(v_1 -v_2)t}\over v_1 e^{(v_1 -v_2)t} -v_2},\;
u_{2}={(v_1 -v_2)v_2 \over v_1 e^{(v_1 -v_2)t} -v_2}, \label{D19c}
\ee
and the Jacobian is:
\be
J= \exp\{ - \int^{t}_{0}\, (u_1 +u_2)(s) \, ds\} = 
{(v_1 -v_2)^{2} e^{(v_1 -v_2)t} \over (v_1 e^{(v_1 -v_2)t} -v_2)^{2}}. 
\label{D19d}
\ee
From (\ref{D19a}) and (\ref{D19d}), we have:
\be
E[u_1u_2] = \int \, dy\, (2\pi D_0 t)^{-N/2}\, 
e^{-|x-y|^2/(2D_0 t)} \, {u_{1}^{0}(y)\, u_{2}^{0}(y) (u_{1}^{0}(y) - 
u_{2}^{0}(y))^{2}\, e^{(u_{1}^{0}(y) - u_{2}^{0}(y))t}\over 
(u_{1}^{0}(y) e^{(u_{1}^{0}(y) - u_{2}^{0}(y))t} -u_{2}^{0}(y))^2}, 
\label{D19e}
\ee
and the total average reaction rate is: 
\be
\int_{R^N}\, E[u_1\,u_2]\, dx\, =  \int_{R^N} \, 
{u_{1}^{0}(y)\, u_{2}^{0}(y) (u_{1}^{0}(y) - 
u_{2}^{0}(y))^{2}\, e^{(u_{1}^{0}(y) - u_{2}^{0}(y))t}\over 
(u_{1}^{0}(y) e^{(u_{1}^{0}(y) - u_{2}^{0}(y))t} -u_{2}^{0}(y))^2}
\, dy. \label{D20}
\ee  

It is natural to compare (\ref{D19e}) and (\ref{D20}) with 
the deterministic case when say $v=0$.
Indeed, we solve two equations: $u_{i,t} = -u_1\, u_2$, to find:
\be
u_{1}={ d u_{1}^{0}(x)\, e^{d t} \over u_{1}^{0}(x) e^{dt} - u_{2}^{0}(x)}, 
\;\; u_{2}={d u_{2}^{0}(x)
 \over u_{1}^{0}(x) e^{dt} - u_{2}^{0}(x)}, \label{D21}
\ee
where $d=d(x)=u_{1}^{0}(x) - u_{2}^{0}(x)$, $u_{i}^{0}\geq 0$, $i=1,2$.   
The total reaction rate is:
\be 
\int_{R^N}\, dx \, {(u_{1}^{0}(x) - u_{2}^{0}(x))^2 u_{1}^{0}(x)u_{2}^{0}(x) 
e^{ (u_{1}^{0}(x) - u_{2}^{0}(x))t} \over 
(u_{1}^{0}(x) e^{ (u_{1}^{0}(x) - u_{2}^{0}(x))t} - u_{2}^{0}(x))^{2}},
\label{D22}
\ee
which agrees exactly with (\ref{D20}) ! This turns out to be true 
more in general on 
a system like (\ref{sem1}). For any smooth incompressible velocity field $v$, 
the integral $\int_{R^n} \; g(u)\, dx$ ($g$ a nonlinear function so that 
$g(u) \in L^1(R^n)$ ) 
is independent of 
$v$ by writing solution in characteristic variables and noting that 
the change of variables from $x$ to the 
characteristic variables has unit Jacobian. However, the local (pointwise)  
average reaction rate $E[u_1u_2]$ decays faster by 
a factor $O(t^{-N/2})$ due to 
advection induced diffusion. 

Similarly, we find the formulas on $E[u_i]$, $i=1,2$:
\ba
E[u_1]& =& \int_{R^N}\,dy\, {u_1^{0}(y) (u_{1}^{0}(y)- u_{2}^{0}(y))
e^{(u_{1}^{0}(y)- u_{2}^{0}(y))t} \over u_{1}^{0}(y)
e^{(u_{1}^{0}(y)- u_{2}^{0}(y))t} - u_{2}^{0}(y)}K(x,y,t), \no \\
E[u_2]& =& \int_{R^N}\,dy\, { (u_{1}^{0}(y)- u_{2}^{0}(y))u_{2}^{0}(y)
 \over u_{1}^{0}(y) e^{(u_{1}^{0}(y)- u_{2}^{0}(y))t} - u_{2}^{0}(y)}K(x,y,t), 
\label{D22a}
\ea
which are convolutions of the deterministic solutions
with zero advection (\ref{D21})
with $K$ the heat kernel of diffusion coefficient $D_0/2$. If the initial 
data $u_1^{0}=\chi (\{x \in R^N: x_{1} > 0\})$, 
$u_{2}^{0}(x)= 1 - u_{1}^{0}(x)$, 
namely the front data, then in the large time limit, 
(\ref{D22a}) behaves like (\ref{F11}) and 
the level curves of $E[u_i]=c$, $i=1,2$, behave like (\ref{F12}). 
In other words, fronts undergo diffusion about the mean location 
$x_1 =0$ in the average sense. 

To summarize, we state:
\begin{prop}Consider the $2\times 2$ model combustion system 
(\ref{sem2}) obeying conditions on the velocity field in Theorem 1;
 and deterministic, nonnegative initial data $(u_1^0,
u_2^0)(x)\in (L^\infty({R}^N))^2$. Then $E[u_1],E[u_2],E[u_1u_2]$ 
are expressed in closed analytical form. The total average reaction rate 
$E[\|u_1u_2\|_{L^1({R}^N)}](t)$ is given by the closed form 
formula (\ref{D22}), the same as the formula when 
the random velocity field is absent. However, $E[u_1u_2](x,t)$ is 
smaller, due to advection induced diffusion, than $u_1u_2(x,t)$ 
in the absence of random advection. The mean concentration fronts 
understood in the sense of level curves of $E[u_1],E[u_2]$ diffuse
about the mean position with diffusion constant $D(0)$.
\end{prop}
We remark that if the mean velocity $\bar{v}$ 
is present, then the PDF equation (\ref{D12}) has on the left hand side 
a term $\bar{v}\cdot \nabla_{x} P$. If $\bar{v}$ is a constant vector, 
this term can be removed by going to the comoving frame $\xi=x - \bar{v}t$, 
and the same results on reaction rates as above can be obtained. 

Now we draw a connection between formula (\ref{D22a}) and 
an empirical procedure in combustion (\cite{Bor}, pp 247-248). 
Let $Z=u_1 -u_2$, then $Z$ satisfy the passive scalar equation 
(without reaction), its PDF denoted by $P(Z)$ is Gaussian with 
kernel $K(x,y,t)$, in view of 
(\ref{D12}). If reaction occurs in a thin zone, that is, 
either $u_1=0$, $u_2\not =0$ or $u_1\not =0$, $u_2=0$, or $u_1=u_2=0$, 
then when $Z < 0$, we have $Z=-u_2$, $u_1=0$; when $Z>0$, $Z=u_1$, $u_2=0$. 
It follows that:
\ba
E[u_1] = \int_{Z>0} ZP(Z)\, dZ, \no \\
E[u_2] = -\int_{Z<0} ZP(Z)\, dZ. \label{D23}
\ea
This provides a way to approximate the average of solutions based on 
the PDF function of the (virtual inert) passive scalar $Z$. 

In case of initial data: 
$u_1^{0}=\chi (\{x \in R^N: x_{1} < 0\})$,
$u_{2}^{0}(x)= 1 - u_{1}^{0}(x)$, 
we are in the thin domain regime. We have from (\ref{D22a}) that:
\[ E[u_1] = \int_{u_{1}^{0} > 0}u_{1}^{0}K(x,y,t) = 
\int_{Z>0} Z P(Z)dZ,\]
similarly for $E[u_2]$, hence recovering the empirical formula (\ref{D23}).  

If the initial data is: $u_{1}^{0}=\chi (\{x \in R^N: x_{1} < 0\})$, 
$u_{2}^{0}=\epsilon$ if $x_1 \in (-\dta, 0)$, $u_{2}^{0} =0$ 
if $x_1 \leq -\dta$, 
$u_{2}^{0}=1$ if $x_1 >0$, we have a reaction zone of thickness $\dta >0$.
Then $E[u_1]$ will differ from the thin reaction zone case above by:
\[ \int_{x_1 \in (-\dta,0)} 
\left ({(u_{1}^{0} - \eps)e^{(u_{1}^{0} -\eps)t}\over 
u_{1}^{0} e^{(u_{1}^{0} -\eps)t} -\eps} -1\right )u_{1}^{0}\, K(x,y,t), \]
which is negative for fixed $t$, and small enough $\eps$. 
The approximation (\ref{D23}) will overestimate $E[u_1]$. 

Now if $u_{2}^{0} = \chi (\{x \in R^N: x_{1} > 0\})$, 
$u_{1}^{0}=1$ if $x_1<0$, $u_{1}^{0}=\eps$, if $x_1 \in (0,\dta)$, 
$u_{1}^{0}=0$, if $x_1 \geq \dta$, then we also have a reaction zone of 
thickness $\dta$. The difference of $E[u_1]$ from the thin 
reaction zone case is:
\[ \int_{x_1 \in (0,\dta)} {\eps (\eps - u_{2}^{0}) e^{(\eps -u_{2}^{0})t}
\over \eps e^{(\eps -u_{2}^{0})t} - u_{2}^{0}} K, \]
which is positive for small $\eps$. Hence the approximation (\ref{D23}) 
will underestimate $E[u_1]$.

\section{Multi-Point PDF Equations and Solutions}
\setcounter{equation}{0}
We derive the multi-point PDF equation by imbedding $n\geq 2$  points 
$x_{1},\cdots,x_n \in R^{N}$ into a vector:
\be
X = (x_1,\cdots, x_n) \in R^{Nn}, \label{M1}
\ee
and corresponding values of solutions into:
\be
\Theta (X,t) = (u(x_1,t),\cdots, u(x_n,t)) \in R^{mn}, \label{M2}
\ee
Let also:
\be
V(X,t) = (v(x_1,t),\cdots, v(x_n,t)), \label{M3}
\ee
and:
\be
F(\Theta) = (f(u(x_1,t)), \cdots, f(u(x_n,t))) \in R^{mn}. \label{M4}
\ee
Then the $n$ equations ($u_i=u(x_i,t)$):
\[ u_{i,t} + v(x_i,t,\omega) \cdot \nabla_{x_i} u_{i} = f(u_{i}), \]
can be written into the system:
\be
\Theta_{t} + V\cdot \nabla_{X} \Theta = F(\Theta), \label{M5}
\ee
which is of the same form as the original equation for $u$. 
It follows that the one point PDF equation of $\Theta$, or 
n-point PDF equation of $u$, is ($P=P(\Theta,X,t)$):
\be
P_{t} + \nabla_{\Theta}\cdot ( (f(u_1),\cdots, f(u_n))P) = 
{1\over 2}\nabla_{X} \cdot (D_n \nabla_{X} P), \label{M6}
\ee
where $D_n$ is the $n\times n$ block matrix $(D(x_k -x_l))$ with 
each block an $N\times N$ covariance matrix of original velocity $v$.
 
We shall need a structure assumption on the original velocity 
covariance matrix $D=D(x)$, $x=(x_1,\cdots,x_N) \in R^N$:
\be
D_{ij}(x) = D(0)\dta_{ij} -d_{ij}(x)
= D(0)\dta_{ij} - D_{1}|x|^{2}\left [ \dta_{ij} 
+{2\over d-1}\left ( \dta_{ij} - {x_i x_j\over |x|^2}\right )\right ],
\label{M7}
\ee
valid for the domain of small $|x|$, also known as the Batchelor regime, 
\cite{Kr68}, \cite{BGK}. Equation (\ref{M6}) can be written as:
\be
P_{t} +\sum_{i=1}^{n}\, \nabla_{u_i}\cdot (f(u_i)P)
= \left [ {D_0\over 2}(\sum_{i=1}^{n} \nabla_{x_i})^{2} 
+ {1\over 2}\sum_{j\not = k}\, d_{\al \beta}(x_j - x_k)
\partial_{x_{j}^{\al}}\partial_{x_{k}^{\beta}}\right ]P\equiv M_n P, \label{M8}
\ee
where $x_{j}^{\al}$ refers to the $\al$-th component of $x_j$, and 
repeated indices mean summation. The operator $M_n$ is degenerate elliptic 
when acting on $L^2$ functions of $x$. If $P$ is translation invariant, 
the first term on the right hand side of (\ref{M8}) vanishes. 

If one neglects reactions, then the last equation is of the same form as
the equation for the Lagrangian $n$-particle transition probability density
$P_n^{t,s}(X;Y)$, which gives the probability to find $n$ Lagrangian particles 
at positions $y_k$ at time $s$ if they were at positions $x_k$ at time $t<s$, 
for $k=1,...,n$. See \cite{G}, p.21. There is a relation between this object
and our multipoint probability density of the scalar amplitudes, $P(\Theta,X,t)$:
\be
P(\Theta;X,t) = \int dY P_n^{0,t}(X;Y) P(\Theta;Y,0). \label{M8.25}
\ee
This relation is, in fact, just the integral solution of our equation (\ref{M8}).
However, it doesn't hold in the general context of spatially non-Lipschitz 
but only H\"{o}lder continuous velocities that was considered in \cite{G}. 
In fact, it holds only in the context of the ``Batchelor regime'' that we consider, 
where velocities are Lipschitz in space. Indeed, a simple consequence of our
equation is that the ensemble-average ``scalar energy'' ${{1}\over{2}}
\int dx<\theta^2(x,t)>$ is conserved in time (without reactions),
whereas this is not true in the case of non-Lipschitz velocities. 

A particular case of interest of equation (\ref{M8}) holds for $n=2$, 
if the velocity field is also rotationally invariant:
\be
P_{t} +\nabla_{u_1}\cdot (f(u_1)P) +\nabla_{u_2}\cdot (f(u_2)P)
= D_2 r^{N-1}(r^{N+1} P_{r})_{r}. \label{M8.5}
\ee
Equation (\ref{M8.5}) is the rigorous justification of a similar 
equation deduced by Kraichnan, (5.10) in \cite{Kr74}. 
He considered a single scalar with no nonlinear reaction. 
His equation 
was written for the probability density $Q(\Delta,k,t)$ of the 2-point 
difference $\Delta=u-v$ of the scalar at wavenumber $k$.
Kraichnan's proposed equation read
\be Q_t = D_2 [ (k(k Q_k)_k + N k Q_k]  \label{Kr74} \ee
in the special case where molecular diffusion vanishes. We obtain 
this equation by changing variables in formula (\ref{M9}) (written 
for the case of $m=1$ and $f\equiv =0$)
from $u$ and $v$ 
to $\Delta$ and $U=(u+v)/2$, by integrating out the second variable $U$,
and by then Fourier transforming from space variable $r$ to wavenumber $k$. 

For the system (\ref{sem2}), if we consider 
 rotationally invariant (isotropic) 
solutions for the two point PDF $P$, then (\ref{M8.5}) becomes:
\be
P_{t}+(-u_1 u_2 P)_{u_1} + (-u_1u_2 P)_{u_2} + 
(-v_1v_2 P)_{v_1} + (-v_1v_2 P)_{v_2} = D_2 r^{N-1}(r^{N+1} P_{r})_{r}, 
\label{M9}
\ee
where $D_2$ is a positive constant; 
we have used $(u_1,u_2)$ to denote the solutions at the first point,  
$(v_1,v_2)$ at the second point, and their separation distance is $r$.     
Assuming that the initial data and solutions decay rapidly 
at space infinities, 
we multiply (\ref{M8}) when $n=2$ by $u_{1}u_{2}v_{1}v_{2}$, integrate over 
$(u,v)$, to get ($E_2 = E[u_1u_2v_1v_2]$ for short):
\be
(E_2)_{t} = D_2 r^{N-1}(r^{N+1} E_{2})_{r}
- \int \,u_1u_2v_1v_2 \, (u_1 +u_2 +v_1 +v_2)\, P(u,v,x,t)\, du\, dv, 
\label{M10}
\ee
implying that $0 \leq E_2 \leq G=G(r,t)$, where $G$ is the 
solution of:
\be
G_{t} = D_2 r^{1-N}(r^{N+1} G_{r})_{r}, \label{M11}
\ee
which is the equation for two point correlation function of 
Kraichnan's passive scalar model in the free decay regime, see 
\cite{Son99}, and \cite{EX}.

To solve either (\ref{M11}) or (\ref{M9}), let us make the change of 
variables: $\xi = N\log r$, then the operator:
\be
D_2 r^{1-N}(r^{N+1} \partial_{r} \cdot )_{r} = {D_2 N^2}
(\partial_{\xi\xi}+\partial_{\xi})\cdot. \label{M12}
\ee
Solution to (\ref{M11}) is then ($D_3 ={D_2 N^2}$):
\ba
G & = & (4\pi D_{3} t)^{-1/2}\int \, d\eta \, 
\exp\{ - (\xi-\eta + D_{3} t)^{2}/(4D_{3}t)\}
G_0(e^{\eta/N}), \no\\
& = & (4\pi D_{3} t)^{-1/2}\int_{0}^{\infty} \, 
\exp\{ - (N\log (r/s)  + D_{3} t)^{2}/(4D_{3} t)\} G_0(s)\, 
Ns^{-1}\, ds, \label{M13}
\ea
which shows exponential decay rate on any compact set of 
$r$ away from zero for fast decaying $G_0(s)$. $G(0,t)$ is however 
conserved due to smoothness of $G$ near $r=0$, or 
not decaying to zero within a neighborhood 
of $O( c_0^{\sqrt{D_{3}t}}\, e^{-D_{3}t/N})$ to leading order, 
$c_0$ a constant $ > 1$. The behavior of $G$ implies that 
$\int_{(x,y) \in \Omega \times \Omega}\, E[u_1(x)u_2(x)u_1(y)u_2(y)]\, dx\, dy
$ converges to zero exponentially as $t\ra \infty$, $\Omega \in R^N$ a 
compact domain. Using $L^2$ decay property of the 
semigroup generated by the operator $M_n$ \cite{JanRai}, 
the decay of all correlator 
functions follow without isotropy. 

We solve (\ref{M9}) in $(u,v,\xi,t)$ by combining Fourier 
transform in $\xi$ and method of characteristics in $(u,v,t)$ as before. 
The result is: 
\ba
  P(u,v,\xi,t) & = & {(u_2 -u_1)^{2}\over (u_2 -u_1 e^{(u_2 -u_1) t})^2} 
{(v_2 -v_1)^{2}\over (v_2 -v_1 e^{(v_2 -v_1) t})^2} 
e^{(u_2 -u_1 + v_2 -v_1)t}\no \\
& \cdot &  
\int_{R^1}\, dy\, (4\pi D_3 t)^{-1/2}\exp\{ 
-|\xi- y + D_{3} t|^2/(4D_3 t)\} \no 
\ea
\[ \cdot  P_0 \left ({ (-u_1 +u_2)\, u_1\, e^{(u_2 -u_1)t}
\over u_2 - u_1 e^{(u_2 -u_1)t}}, 
{(u_2 -u_1)u_2 \over u_2 - u_1 e^{(u_2 -u_1)t}}, 
{ (-v_1 +v_2)\, v_1\, e^{(v_2 -v_1)t}
\over v_2 - v_1 e^{(v_2 -v_1)t}}, 
{(v_2 -v_1)u_2 \over v_2 - v_1 e^{(v_2 -v_1)t}},e^{y/N}\right ),\]
where $P_0=P_0(u,v,r)$ is the initial PDF data. 
Back to the original variable $r$, we have the two-point PDF formula:
\ba
& &  P(u,v,r,t) = {(u_2 -u_1)^{2}\over (u_2 -u_1 e^{(u_2 -u_1) t})^2} 
{(v_2 -v_1)^{2}\over (v_2 -v_1 e^{(v_2 -v_1) t})^2} 
e^{(u_2 -u_1 + v_2 -v_1)t}\no \\
& \cdot & \int_{0}^{\infty}\, N\,s^{-1}\, ds\,
  (4\pi D_3 t)^{-1/2}\exp\{ -|N\log (r/s) + D_{3} t|^2/(4D_3 t)\} \no \\  
& & P_0 \left ({ (-u_1 +u_2)\, u_1\, e^{(u_2 -u_1)t}
\over u_2 - u_1 e^{(u_2 -u_1)t}}, 
{(u_2 -u_1)u_2 \over u_2 - u_1 e^{(u_2 -u_1)t}}, 
{ (-v_1 +v_2)\, v_1\, e^{(v_2 -v_1)t}
\over v_2 - v_1 e^{(v_2 -v_1)t}}, 
{(v_2 -v_1)u_2 \over v_2 - v_1 e^{(v_2 -v_1)t}},s\right ). \no \\
\label{M15}
\ea

Summarizing the results of this section, we state:
\begin{theo}
Let $P(u_1,...,u_n,x,t)$ be the $n$-point probability density function
of the solution $u(x,t)$ of system (\ref{sem1}) with advection velocity
satisfying the conditions of Theorem 1. Then $P$ obeys the closed 
equation (\ref{M6}). If the separations $r_{ij}=|x_i-x_j|$ are all small
for $i,j=1,...,n$, then the equation takes the special form (\ref{M8}).
If in addition $v$ is rotationally invariant, the 2-point PDF 
$P(u_1,u_2,x,t)$ obeys the simplified equation (\ref{M9}) 
and the closed form analytical 
solution for $P$ is given by (\ref{M15}) for the case of deterministic
initial data of the system (\ref{sem2}).
\end{theo}

\section{Concluding Remarks}
Under the white noise assumption of incompressible advection velocity field, 
PDF equations are closed for the solution of the 
semilinear hyperbolic system (\ref{sem1}).   
Exact formulas for one-point and multi-point PDFs (under isotropy) 
on solutions of the isothermal non-premixed turbulent flame system 
(\ref{sem2}) can be derived. From exact formulas, one is able to 
recover empirical formulas in combustion on 
averaged solutions in the thin reaction zone limit, and analyze 
the effect of finite domain zones on the empirical formulas. 
It is found that they can either overestimate or underestimate 
averaged solutions.
It is also found that fronts understood as level 
curves of averaged concentration variables, diffuse about the 
mean position in the ensemble average sense. The front diffusion about 
mean location is known in the distribution sense in the context of 
the Burgers' equation and other one dimensional convex scalar 
conservation laws, see \cite{Xin0}. It is conceivable that 
this picture is valid for turbulent combustion fronts when 
the white noise assumption is relaxed. A future study with 
assistance of numerical computation appears feasible.  
It will also be interesting to study more geometric 
properties of random level curves 
using higher order PDFs, such as its roughness and dimensions in 
the large time limit for front data in several space dimensions. 
See \cite{Cons94} for results  
in this direction with a different approach. 

\section{Acknowledgements}
G. E. would like to thank the DOE for partial support of this work 
through grant LDRD-ER 2000047 at the Los Alamos National Laboratory. 
Work of J. X. was partially supported by NSF and ARO, 
and the visiting professorship 
at the Research Institute of Electronic Sciences of Hokkaido University 
during his visit there in June of 2000. J. X. would especially 
like to thank his host Prof. Yasumasa Nishiura for his hospitality and for 
many helpful discussions.

\end{document}